\newread\epsffilein    
\newif\ifepsffileok    
\newif\ifepsfbbfound   
\newif\ifepsfverbose   
\newdimen\epsfxsize    
\newdimen\epsfysize    
\newdimen\epsftsize    
\newdimen\epsfrsize    
\newdimen\epsftmp      
\newdimen\pspoints     
\pspoints=1bp          
\epsfxsize=0pt         
\epsfysize=0pt         
\def\epsfbox#1{\global\def\epsfllx{72}\global\def\epsflly{72}%
   \global\def\epsfurx{540}\global\def\epsfury{720}%
   \def\lbracket{[}\def\testit{#1}\ifx\testit\lbracket
   \let\next=\epsfgetlitbb\else\let\next=\epsfnormal\fi\next{#1}}%
\def\epsfgetlitbb#1#2 #3 #4 #5]#6{\epsfgrab #2 #3 #4 #5 .\\%
   \epsfsetgraph{#6}}%
\def\epsfnormal#1{\epsfgetbb{#1}\epsfsetgraph{#1}}%
\def\epsfgetbb#1{%
%
%
\openin\epsffilein=#1
\ifeof\epsffilein\errmessage{I couldn't open #1, will ignore it}\else
%
%
   {\epsffileoktrue \chardef\other=12
    \def\do##1{\catcode`##1=\other}\dospecials \catcode`\ =10
    \loop
       \read\epsffilein to \epsffileline
       \ifeof\epsffilein\epsffileokfalse\else
%
%
          \expandafter\epsfaux\epsffileline:. \\%
       \fi
   \ifepsffileok\repeat
   \ifepsfbbfound\else
    \ifepsfverbose\message{No bounding box comment in #1; using defaults}\fi\fi
   }\closein\epsffilein\fi}%
%
%
\def\epsfclipstring{}
\def\epsfsetgraph#1{%
   \epsfrsize=\epsfury\pspoints
   \advance\epsfrsize by-\epsflly\pspoints
   \epsftsize=\epsfurx\pspoints
   \advance\epsftsize by-\epsfllx\pspoints
%
%
   \epsfxsize\epsfsize\epsftsize\epsfrsize
   \ifnum\epsfxsize=0 \ifnum\epsfysize=0
      \epsfxsize=\epsftsize \epsfysize=\epsfrsize
      \epsfrsize=0pt
%
%
     \else\epsftmp=\epsftsize \divide\epsftmp\epsfrsize
       \epsfxsize=\epsfysize \multiply\epsfxsize\epsftmp
       \multiply\epsftmp\epsfrsize \advance\epsftsize-\epsftmp
       \epsftmp=\epsfysize
       \loop \advance\epsftsize\epsftsize \divide\epsftmp 2
       \ifnum\epsftmp>0
          \ifnum\epsftsize<\epsfrsize\else
             \advance\epsftsize-\epsfrsize \advance\epsfxsize\epsftmp \fi
       \repeat
       \epsfrsize=0pt
     \fi
   \else \ifnum\epsfysize=0
     \epsftmp=\epsfrsize \divide\epsftmp\epsftsize
     \epsfysize=\epsfxsize \multiply\epsfysize\epsftmp   
     \multiply\epsftmp\epsftsize \advance\epsfrsize-\epsftmp
     \epsftmp=\epsfxsize
     \loop \advance\epsfrsize\epsfrsize \divide\epsftmp 2
     \ifnum\epsftmp>0
        \ifnum\epsfrsize<\epsftsize\else
           \advance\epsfrsize-\epsftsize \advance\epsfysize\epsftmp \fi
     \repeat
     \epsfrsize=0pt
    \else
     \epsfrsize=\epsfysize
    \fi
   \fi
%
%
   \ifepsfverbose\message{#1: width=\the\epsfxsize, height=\the\epsfysize}\fi
   \epsftmp=10\epsfxsize \divide\epsftmp\pspoints
   \vbox to\epsfysize{\vfil\hbox to\epsfxsize{%
      \ifnum\epsfrsize=0\relax
        \includegraphics{#1}%
      \else
        \epsfrsize=10\epsfysize \divide\epsfrsize\pspoints
        \includegraphics{#1}%
      \fi
      \hfil}}%
\global\epsfxsize=0pt\global\epsfysize=0pt}%
%
%
{\catcode`\%=12 \global\let\epsfpercent=
%
%
\long\def\epsfaux#1#2:#3\\{\ifx#1\epsfpercent
   \def\testit{#2}\ifx\testit\epsfbblit
      \epsfgrab #3 . . . \\%
      \epsffileokfalse
      \global\epsfbbfoundtrue
   \fi\else\ifx#1\par\else\epsffileokfalse\fi\fi}%
%
%
\def\epsfempty{}%
\def\epsfgrab #1 #2 #3 #4 #5\\{%
\global\def\epsfllx{#1}\ifx\epsfllx\epsfempty
      \epsfgrab #2 #3 #4 #5 .\\\else
   \global\def\epsflly{#2}%
   \global\def\epsfurx{#3}\global\def\epsfury{#4}\fi}%
%
%
\def\epsfsize#1#2{\epsfxsize}
%
%

\input harvmac
\def\figflag{I}
\def\tfig#1{Fig.~\the\figno\xdef#1{Fig.~\the\figno}\global\advance\figno by1}
\def\figI{I}
%
\newdimen\tempszb \newdimen\tempszc \newdimen\tempszd \newdimen\tempsze
\ifx\figflag\figI
%
\def\epsfsize#1#2{\expandafter\epsfxsize{
 \tempszb=#1 \tempszd=#2 \tempsze=\epsfxsize
     \tempszc=\tempszb \divide\tempszc\tempszd
     \tempsze=\epsfysize \multiply\tempsze\tempszc
     \multiply\tempszc\tempszd \advance\tempszb-\tempszc
     \tempszc=\epsfysize
     \loop \advance\tempszb\tempszb \divide\tempszc 2
     \ifnum\tempszc>0
        \ifnum\tempszb<\tempszd\else
           \advance\tempszb-\tempszd \advance\tempsze\tempszc \fi
     \repeat
\ifnum\tempsze>\hsize\global\epsfxsize=\hsize\global\epsfysize=0pt\else\fi}}
\epsfverbosetrue
\fi
%
%
%
%
%

\def\ifigure#1#2#3#4{
\midinsert
\vbox to #4truein{\ifx\figflag\figI
\vfil\centerline{\epsfysize=#4truein\epsfbox{#3}}\fi}
\baselineskip=12pt
\narrower\narrower\noindent{\bf #1:} #2
\endinsert
}
%
%
\def\ifigures#1#2#3#4#5#6#7#8{
\midinsert
\centerline{
\hbox{\vbox{
\divide\hsize by 2
\vbox to #4truein{\ifx\figflag\figI
\vfil\centerline{\epsfysize=#4truein\epsfbox{#3}}\fi}
\baselineskip=12pt
\narrower\narrower\noindent{\bf #1:} #2
}}\qquad
\hbox{\vbox{
\divide\hsize by 2
\vbox to #8truein{\ifx\figflag\figI
\vfil\centerline{\epsfysize=#8truein\epsfbox{#7}}\fi}
\baselineskip=12pt
\narrower\narrower\noindent{\bf #5:} #6
}}}
\endinsert
}

\lref\collective{W.\ Fischler, S.\ Paban, and M.\ Rozali, ``Collective
Coordinates for D-branes'' \hfil\break 
[hep-th/9604014], to be published in {\it Phys.\ Lett.}~{\bf B}.}

\lref\scattrefs{I.\ R.\ Klebanov and L.\ Thorlacius, {\it Phys.\
Lett.}~{\bf B371}, 51 (1996)\semi
C.\ Bachas, ``D-Brane Dynamics'' [hep-th/9511043]\semi
M.\ R.\ Garousi and R.\ C.\ Myers, ``Superstring Scattering from
D-Branes'' \hfil\break
[hep-th/9603194].}

\lref\dbrane{J.\ Dai, R.\ Leigh, and J.\ Polchinski, {\it Mod.\ Phys.\
Lett.}~{\bf A4}, 2073 (1989)\semi 
J.\ Polchinski, S.\ Chaudhuri, and  C.\ V.\ Johnson, ``Notes on
D-Branes'' \hfil\break
[hep-th/9602052] and references therein.} 

\lref\logop{A.\ B.\ Zamolodchikov, {\it JETP Lett.}~{\bf 43}, 730
(1986); {\it Sov.\ J.\ Nuc.\ Phys.}~{\bf 46}, 1090 (1987)\semi
I.\ I.\ Kogan and N.\ E.\ Mavromatos, ``World Sheet
Logarithmic Operators and Target Space Symmetries in String Theory'' 
[hep-th/9512210].}

\lref\bogo{The relationship between the mass of solitons and the
norm of the zero-mode was noted in E.\ B.\
Bogomol'nyi, {\it Sov.\ J.\ Nuc.\ Phys.}~{\bf 24}, 449 (1976).}

\lref\forces{C.\ G.\ Callan and I.\ G.\ Klebanov, ``D-Brane Boundary
State Dynamics'' \hfil\break [hep-th/9511173]\semi 
T.\ Banks and L.\ Susskind, ``Brane-Anti-Brane Forces'' 
[hep-th/9511194]\semi
U.\ H.\ Danielsson, G.\ Ferretti, and B.\ Sundborg, ``D Particle
Dynamics and Bound States'' [hep-th/9603081]\semi
D.\ Kabat and P.\ Pouliot, ``A Comment on Zero-brane Quantum
Mechanics''\hfil\break [hep-th/9603127]. }

\Title{UTTG-06-96}{Virtual D-Branes} 

\centerline{David Berenstein, Richard Corrado, Willy Fischler,
Sonia Paban, and Moshe Rozali\footnote{*}{Research supported in part
by the Robert A.\ Welch Foundation and NSF Grant PHY~9511632.}}
\bigskip\centerline{\it Theory Group and Department of Physics}
\centerline{\it University of Texas at Austin, Austin, TX 78712}

\vskip .3in

Using a formalism developed to include collective coordinates, we
calculate the contributions to S-matrix elements due to off-shell
D-branes in a string coupling expansion. The formalism is further used
to establish a two-dimensional computation of higher order corrections
to the D-brane tension, both for the bosonic and the supersymmetric cases.

\Date{22 May 1996} 

\newsec{Introduction}

In this paper, we apply the collective coordinate formalism developed
in an earlier work~\collective\ to include higher order corrections
for scattering off D-branes~\scattrefs. One goal is to explicitly show how
off-shell D-branes appear as intermediate states in such amplitudes.  
This constitutes a check that the formalism properly accounts for
the brane's recoil at the quantum mechanical level. One does indeed 
expect that virtually recoiling D-branes should be properly included
in the formalism. After all, the translational zero modes to which
these collective coordinates are related already appear in the
world-sheet formulation as fully quantized massless open string states.

In the first part of the paper, we show how the formalism includes the
contribution from off-shell recoiling branes by considering the
scattering of a closed string state off a Dirichlet 0-brane. In the
last part, we establish a formula for the corrections to the D-brane
tension~\dbrane\ and calculate them in the bosonic case using a conformal
field theory approach. This two-dimensional formula is then used to
illustrate, to this order, the expected non-renormalization of the
mass of BPS saturated branes in the supersymmetric case.

\newsec{Virtual Recoil}

For the purpose of illustrating what to expect from virtually
recoiling heavy branes, we first consider the simple case of a light
scalar field (representing a light closed string state, for example
the closed string tachyon) scattering off a heavy particle
(representing the 0-brane). The coupling of the light particle to the
heavy one, $\lambda$, is taken to be independent of the string
coupling constant. This mimics faithfully the coupling of the closed
string tachyon to the D-brane. This coupling can be calculated, at tree
level, as the one-point closed string tachyon amplitude on the disc,
thereby obtaining 
$$\lambda_0={1\over 2\pi}.$$

To evaluate the scattering amplitude of a heavy particle by a light
one, it is helpful to use ``old-fashioned'' perturbation theory. The
initial state for the process consists of a heavy particle (of
mass~$M$) at rest ($\vec{p}=0$) and a light particle with
energy~$\omega_1$ and momentum~$\vec{k}_1$. The final state has a
light particle with energy~$\omega_2$ and momentum~$\vec{k}_2$ and
a heavy particle with momentum~$\vec{k}_1-\vec{k}_2$ and
energy~$M+(\vec{k}_1-\vec{k}_2)^2/2M$. At each interaction vertex,
momentum is conserved, whereas energy is only conserved between the
initial and final states. This implies
$$M+\omega_1=M+{(\vec{k}_1-\vec{k}_2)^2\over 2M} +\omega_2.$$

Since our purpose is to focus on the contribution from virtually
recoiling branes, we need to consider, in this example, two 
diagrams\footnote{$^\dagger$}{There are also contributions from contact
interactions, but, to this order in string coupling, the quartic
interaction need only be used to first order in perturbation theory.}
at the lowest relevant order in perturbation theory, shown 
in~\tfig\onepinterm\ and~\tfig\threepinterm. The first diagram 
contains a one particle intermediate state: a heavy particle with 
energy~$M+\vec{k}_1^2/2M$ and momentum~$\vec{k}_1$. The second
diagram contains a three particle intermediate state: a heavy particle
with energy~$M+\vec{k}_2^2/2M$ and momentum~$-\vec{k}_2$ and two light
particles with energies~$\omega_1,\omega_2$ and 
momenta~$\vec{k}_1,\vec{k}_2$, respectively.
\ifigures\onepinterm{The diagram with a one particle intermediate
state.}{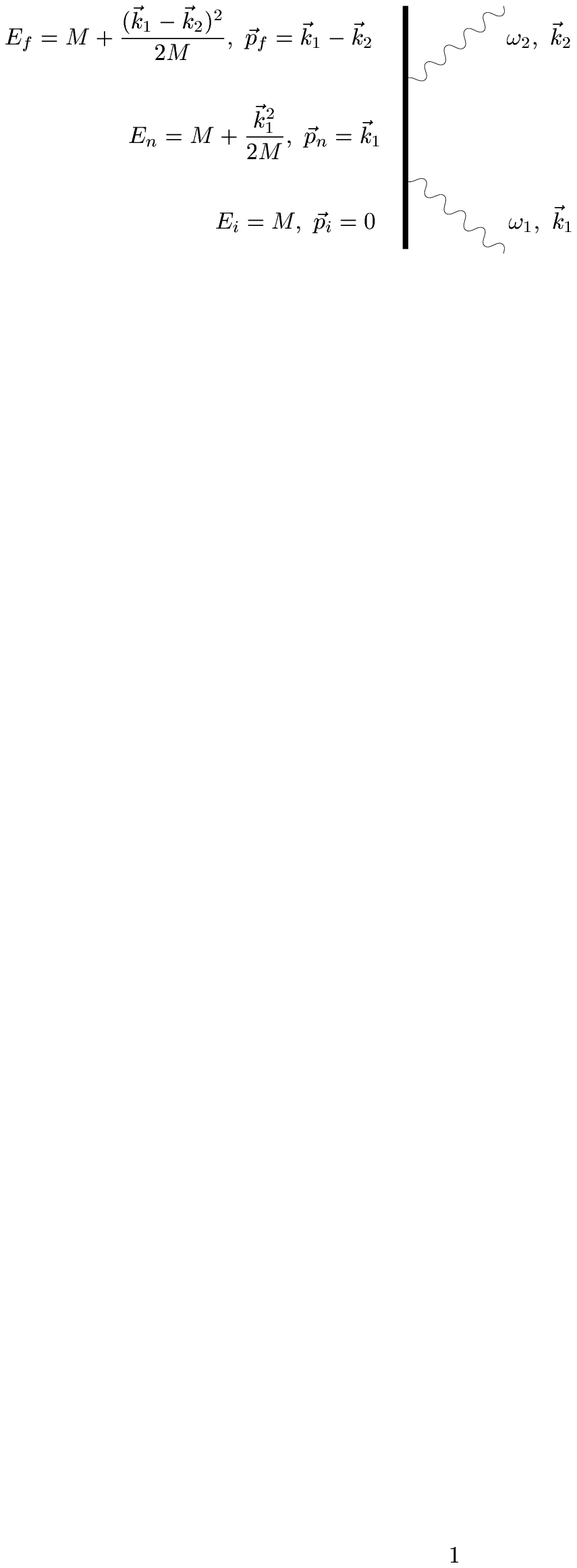}{2}\threepinterm{The diagram with a three
particle intermediate state.}{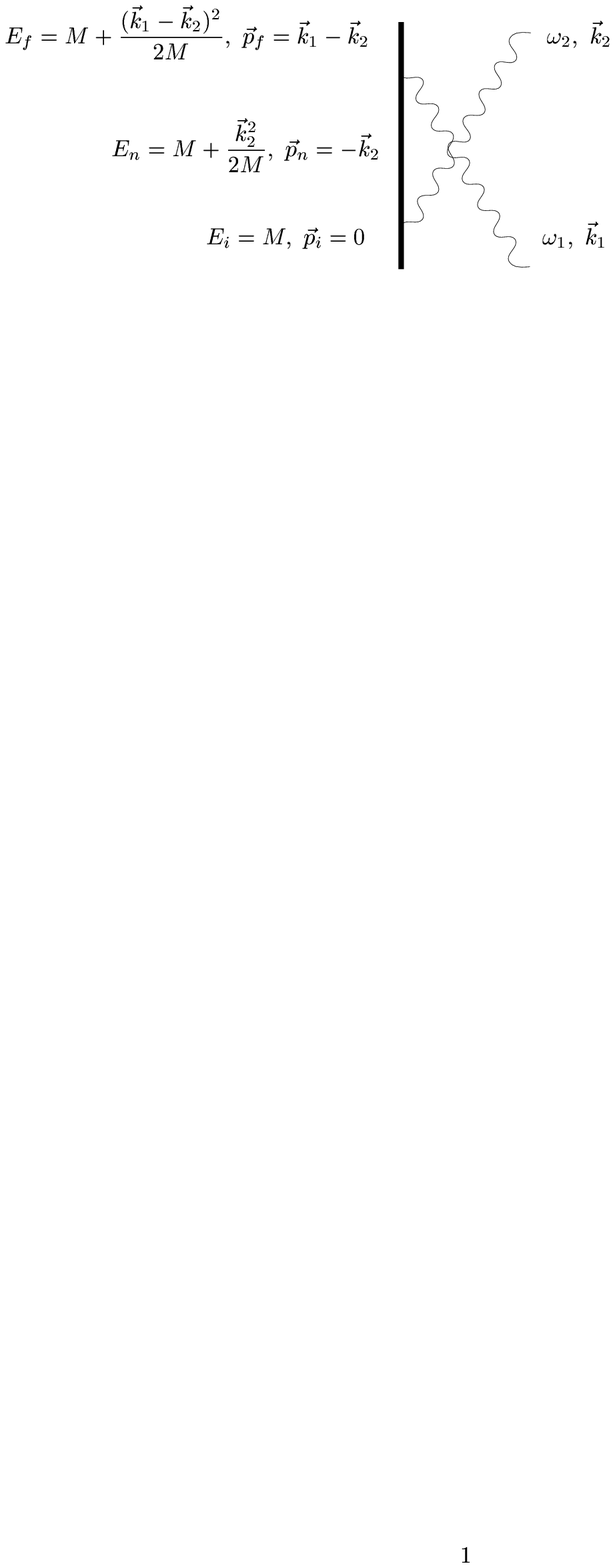}{2}
The amplitude corresponding to \onepinterm\ is 
$$A_1={\lambda^2\over (M+\omega_1)-(M+{\vec{k}_1^2\over 2M})}
={\lambda^2\over \omega_1-{\vec{k}_1^2\over 2M}},$$
while that for \threepinterm\ is
$$A_2={\lambda^2\over (M+\omega_1)-(M+{\vec{k}_2^2\over 2M}
+\omega_1+\omega_2)}
= {-\lambda^2\over \omega_2+{\vec{k}_2^2\over 2M}}.$$
Combining these contributions to the scattering amplitude that involve a
virtually recoiling particle gives
\eqn\ampvrhp{A={\lambda^2\over M} 
{\vec{k}_1\cdot\vec{k}_2\over \omega_1^2}
\left(1- {\vec{k}_1\cdot(\vec{k}_1-\vec{k}_2)\over M\omega_1}\right)^{-1}.}
The leading term, since $M$ is taken to be of order $g^{-1}$, is
\eqn\amplead{A_0={\lambda^2\over M} {\vec{k}_1\cdot\vec{k}_2\over
\omega_1^2}.}
This expression agrees, to the same order in~$g$, with the result
obtained in~\collective\ for the S-matrix element associated to a
closed string tachyon scattering off a 0-brane when factorizing onto
intermediate translational zero-modes.  It is then
simple to read off the D-brane's mass, $M$, from~\amplead\ and
find that it agrees with that obtained previously
in~\refs{\dbrane,\collective}. 
This way of identifying the mass is well suited for calculating higher
order corrections to the brane's mass, as is shown in section~3.  

The contributions arising from the heavy particle being off-shell
first appear in the subleading terms of equation~\ampvrhp. We now show
how the formalism developed in~\collective\ reproduces these terms. 
As was explained in~\collective, conformal invariance requires, in
particular, the presence in the two dimensional action of the
coupling\footnote{$^\ddagger$}{Note that this is a logarithmic
operator~\logop.}  
\eqn\confreq{{1\over 8\pi}\oint ds\, \vec{a}(X_0)\cdot\partial_n\vec{X},}
with $X_0$ being target time and 
$$\vec{a}(X_0)={(\vec{k}_1-\vec{k}_2)X_0\over M}.$$
The inclusion of this term in the action modifies the amplitude for
the scattering of a closed string tachyon off the D-brane
\eqn\twotach{S= \left\langle g\int d^2z_1\, e^{i k_1\cdot X(z_1)}
\, g \int d^2z_2\, e^{-i k_2\cdot X(z_2)} \right\rangle_{\rm Disc}.}
This amplitude factorizes onto intermediate open string states in the
limit where one of the vertex operators approaches the
boundary. Including the new term~\confreq, we obtain in this limit
\eqn\discfact{S=g\int_0^1 rdr\,
r^{2(\omega_1\omega_2-\vec{k}_1\cdot\vec{k}_2)}
(1-r^2)^{-2-2\omega_1^2} 
(1-r^2)^{2\omega_1\vec{k}_1\cdot(\vec{k}_1-\vec{k}_2)/M}.}

From~\discfact, one observes that all open string poles are shifted from
their lowest order values. In particular, the contribution from the
translational zero mode is modified 
$${1\over \omega_1^2} \longrightarrow 
{1\over \omega_1^2- \omega_1\vec{k}_1\cdot(\vec{k}_1-\vec{k}_2)/M}.$$
This shift reproduces the earlier result~\ampvrhp, which was obtained
by a simple quantum mechanical derivation. This, then, shows that the
formalism developed in~\collective\ properly accounts for the
contributions from off-shell
D-branes\footnote{$^\diamondsuit$}{From~\discfact\ one can also recover the
contact interactions mentioned in a previous footnote.}.

\newsec{Renormalization of the D-brane Tension}

In order to set up the calculation of the tension renormalization, we
first study the bosonic case, where such corrections are 
non-vanishing. The notorious singularities associated to open and
closed string tachyons that plague bosonic string theory can be
easily isolated and do not affect the formulation of higher order
corrections to the D-brane mass.

We start by reminding the reader that the mass of the brane was
obtained earlier (in equation~\amplead) by factorizing onto the
translational zero-mode poles. This suggests a simple generalization to
account for higher order corrections. Consider the scattering
amplitude involving closed string tachyons described in
equation~\twotach, calculated now on the annulus. By
letting the two vertex operators approach either boundary of the
annulus, as in~\tfig\annfact, the amplitude factorizes onto
intermediate open string states. 
\vskip3mm
\ifigure\annfact{Factorization on the annulus.}{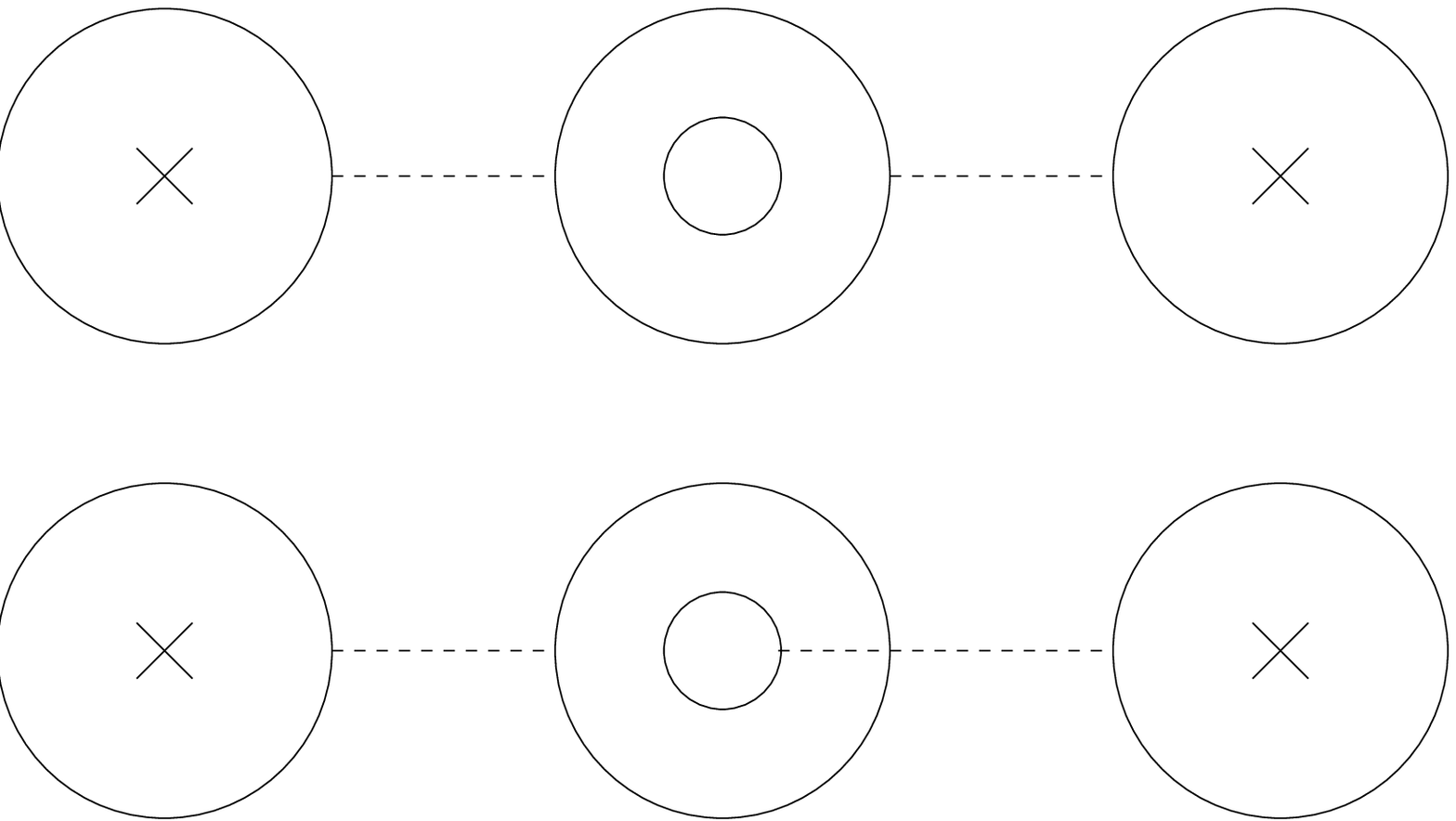}{1.2}
In the limit depicted in~\annfact, one can identify the one-loop
correction to the translational zero-mode propagator
$${1\over \omega^2}\longrightarrow {1\over \omega^2+\Sigma(\omega^2)},$$
where 
\eqn\selfenergy{\Sigma(\omega^2) = {M_0 g^2\over 64\lambda_0^2}
\left\langle \oint ds_1 \, e^{i\omega X_0(s_1)}\partial_n\vec{X}(s_1)
\oint ds_2 \, e^{-i\omega X_0(s_2)}\partial_n\vec{X}(s_2) \right\rangle}
and~$M_0$ and~$\lambda_0$ are the tree level mass and coupling,
respectively. To find the mass correction to the D-brane, we need to
extract the contributions to the amplitude arising from intermediate
translational zero-modes, of the
form~$\vec{k}_1\cdot\vec{k}_2/\omega^2$, as in equation~\amplead. The
one-loop corrected amplitude, $A$, obtained in this way is
$A=A_0+\Delta A$, where 
\eqn\loopamp{\Delta A=-{\lambda_0^2\over M_0^2}
{\vec{k}_1\cdot\vec{k}_2\over\omega^2}\Sigma^\prime(0) .}
One recognizes from this expression that the mass 
correction, $\Delta M$, corresponds to the wavefunction renormalization
of the translational zero-modes~\bogo; we find
\eqn\masscorr{\Delta M = M_0 \Sigma^\prime(0).} 

For this approach to be meaningful, there should be no double pole
at~$\omega^2=0$. Indeed, the presence of such a double pole would
correspond to generating a mass for the translational zero-modes,
which in turn implies that the D-brane mass is position dependent. The
residue of this double pole is proportional to the two-point function
for zero modes on the annulus
\eqn\dpres{\Sigma(0) \propto \left\langle \oint ds_1\, 
\partial_n\vec{X}(s_1) \oint ds_2\, \partial_n\vec{X}(s_2) 
\right\rangle_{\rm Annulus},}
where the line integrals are evaluated on the boundaries of the
annulus implied in~\annfact. The various terms cancel among
themselves, leaving one term related to the existence, at this order,
of an open string tachyon tadpole, depicted in~\tfig\tachtadpole. Such
a term is not present in the supersymmetric case, so we consider it
merely a symptom of bosonic string pathology. 
\ifigure\tachtadpole{Contribution due to the one-loop open string
tachyon tadpole.}{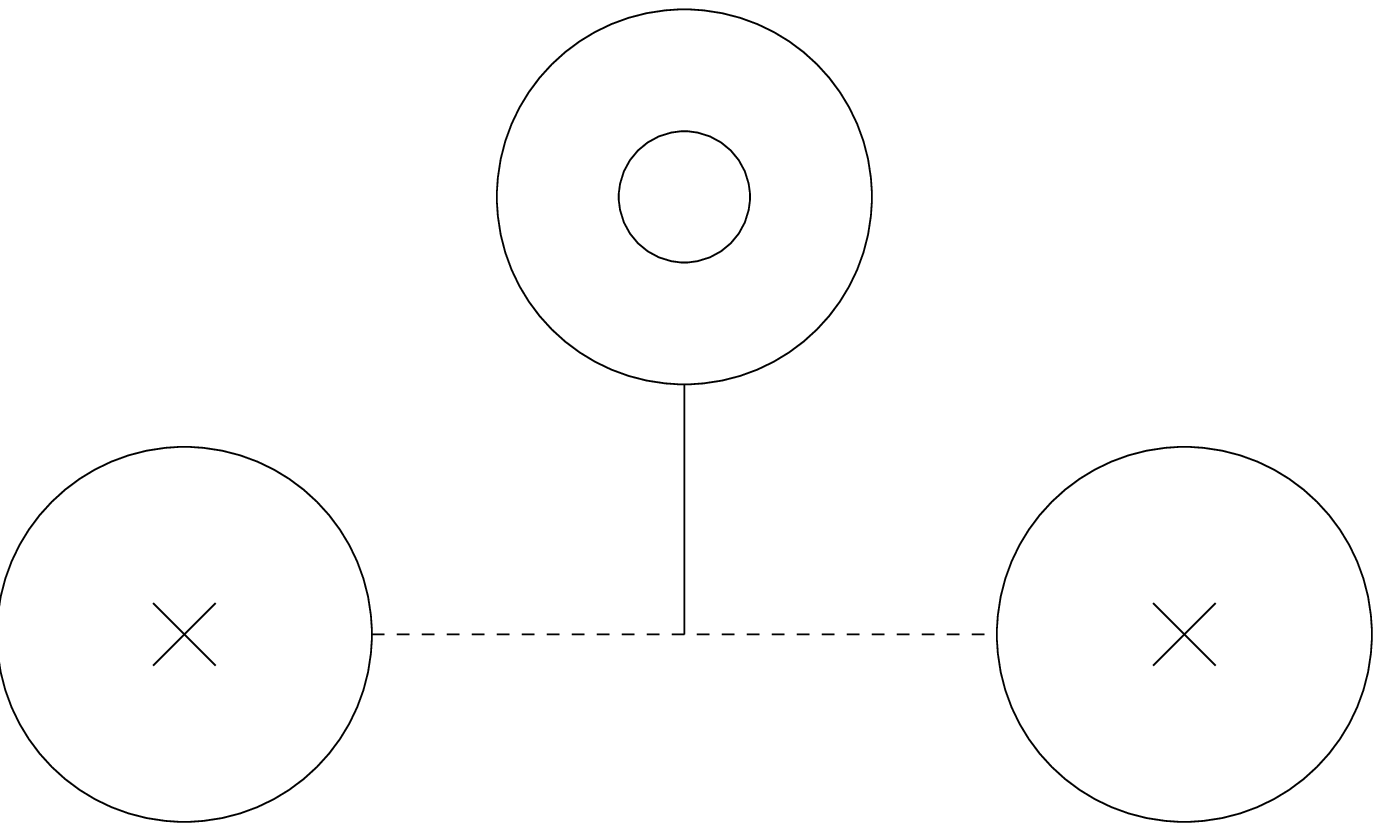}{1}
This result, namely that the double pole receives a contribution
solely from the presence of an open string tachyon tadpole, is easily
obtained by using the analytic properties of the Dirichlet Green's
function on the annulus. It is most convenient to parameterize the
annulus as a finite strip in the upper half plane, with the opposite
ends identified, as in~\tfig\strip.   
\ifigure\strip{Parameterization of the annulus by a strip.}{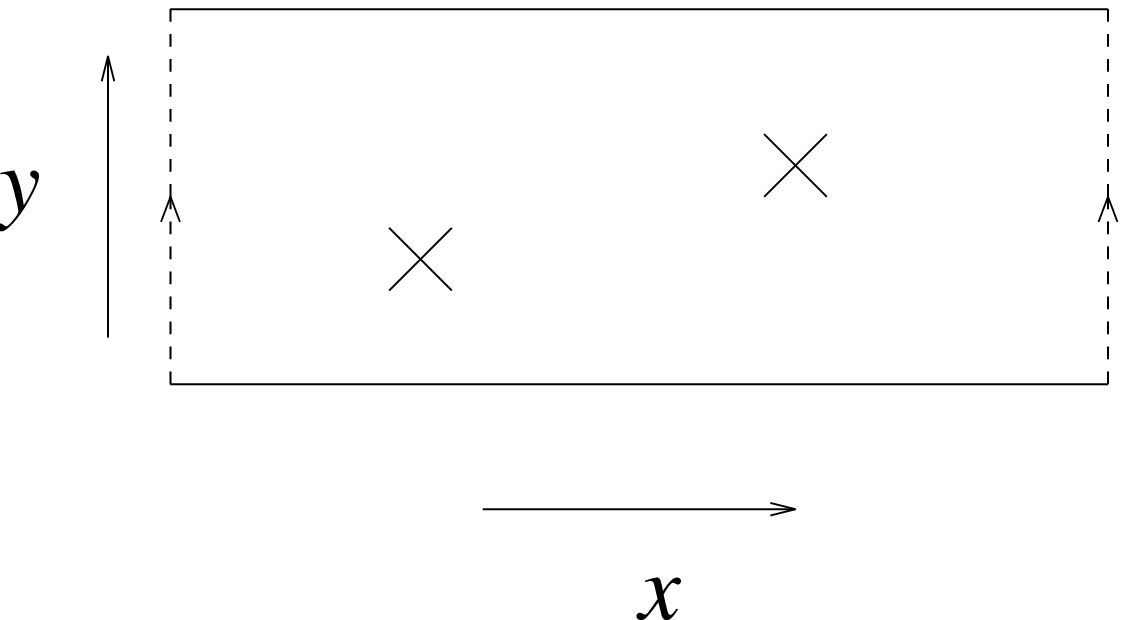}{1.3}
The double pole residue calculated from~\dpres\ takes the explicit
form 
\eqn\dpexplicit{\Sigma(0)\propto  \int {da\over a^3}
\left[\prod_{n=1}^\infty(1-a^{2n})\right]^{-24} \oint ds_1 \oint ds_2 \,  
\partial_n\partial_n G_D (s_1,s_2),}
where $a$ is the modulus of the annulus and the integrals are over
both boundaries. Using Gauss' theorem and introducing a cutoff on the
integrals, one can 
deform the contour of integration of one of the vertex operators to a
semicircle of radius~$\epsilon$, shown in~\tfig\gauss. In this way, we
isolate the divergence due to the operators becoming close and find that
there is a ~$1/\epsilon$ cutoff dependence. This corresponds to
the open string tachyon tadpole, as indicated in~\tachtadpole. 
\ifigure\gauss{The deformation of contours leading to the tachyon
tadpole.}{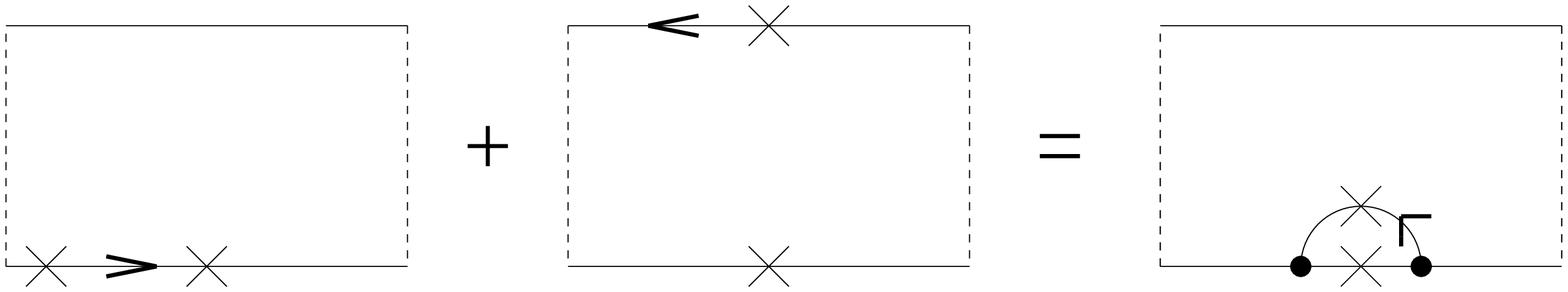}{1} 
We next turn to the evaluation of the mass correction by calculating
the residue of the single pole in~$\omega^2$,
$\Sigma^\prime(0)$. Equations~\selfenergy\ and~\masscorr\ enable us to
identify the mass correction
\eqn\idmass{\Delta M = \left. {g^2 M^2_0\over 64\lambda_0^2}
{\partial\over\partial(\omega^2)} \right|_{\omega=0}
\left\langle\oint ds_1\, e^{i\omega X_0(s_1)} \partial_n\vec{X}(s_1) 
\oint ds_2\, e^{-i\omega X_0(s_2)} \partial_n\vec{X}(s_2) 
\right\rangle_{\rm Annulus}.}
Disregarding contributions which are proportional to open string
tachyon tadpoles, we obtain 
\eqn\finitem{\Delta M = -{1\over 2\pi}\int_0^1 {da\over a^3}
\left[\prod_{n=1}^\infty (1-a^{2n})\right]^{-24},}
which is simply the one-loop free energy.

We are now in a position to apply this approach to calculating mass
corrections in the supersymmetric case. For the sake of clarity, we
focus on the 0-branes of Type~IIA string theory. Our approach can be
easily generalized to higher $p$-branes by suitably compactifying the
appropriate number of Neumann directions, thereby making the mass of
the brane finite. In what follows, we show that the mass corrections
vanish as one sums over different spin structures. This result is
expected, since D-branes are BPS saturated states, and should
be considered as a warm-up for calculating finite
corrections to physical quantities involving non-BPS saturated states. 

To calculate mass corrections in the supersymmetric case, we consider
the supersymmetric extension of~\selfenergy
\eqn\supercase{\eqalign{\Sigma(\omega^2)\propto &\left\langle  \oint ds_1 \,
e^{i\omega X_0(s_1)} \left(\partial_n\vec{X}(s_1) + i\omega
\psi_0(s_1)\vec{\psi}(s_1)\right) \right.\cr
&~~\times \left. \oint ds_2 \, e^{-i\omega X_0(s_2)}
\left(\partial_n\vec{X}(s_2) - i\omega  
\psi_0(s_2)\vec{\psi}(s_2)\right) \right\rangle .\cr}}
The addition over spin structures factors out of the calculation for
the correlation functions involving worldsheet bosons
$$\left\langle \oint ds_1\, e^{i\omega X_0(s_1)}\partial_n\vec{X}(s_1)\,
\oint ds_2 \, e^{-i\omega X_0(s_2)}\partial_n\vec{X}(s_2) \right\rangle.$$  
As is well known, the sum over spin structures vanishes by the
abstruse identity. Therefore both the single and double pole residues
receive no contribution from worldsheet bosons, thus 
\eqn\fermcorr{\Sigma^\prime(0)\propto \left\langle \oint ds_1\,
\psi_0(s_1)\vec{\psi}(s_1) \oint ds_2\, 
\psi_0(s_2)\vec{\psi}(s_2)\right\rangle}
and $\Sigma(0)=0$. 
To further evaluate $\Sigma^\prime(0)$, note that contributions from 
worldsheet fermion correlation functions are periodic regardless of
the spin structure. Therefore, one can again
analytically deform the integration contour of one of the vertex
operators, as in~\gauss. This once again only results in a
contribution which is
proportional to a~$1/\epsilon$ cutoff dependence. In this case, since
the result is the same for all spin structures, it vanishes by the
abstruse identity.

\newsec{Conclusions}

D-branes are an important ingredient in the various dualities recently
discovered in string theory. In the weak coupling limit, these
solitons are heavy and can therefore be described reasonably well by
classical mechanics. We have shown how, in the first quantized
approach to string theory ({\it i.e.}, the sum over worldsheets),
off-shell D-branes are included and contribute to scattering matrix
elements. This is a preliminary step in developing the quantum
mechanics of D-branes. In order to obtain these results, it was
crucial to introduce collective coordinates for the D-brane. 

Collective coordinates, needed for the study of the dynamics of
D-branes, also provide a means of calculating higher order corrections
to the brane's tension. For the superstring, this allowed us to
recover, using a worldsheet approach, the well known
non-renormalization of the D-brane tension. One can envisage other
physical quantities to which higher order corrections are
nonvanishing. It should be possible, for example, to calculate the
corrections to the interaction potential between branes that are
moving with respect to each other~\forces.  

In summary, a better understanding of the dynamics of D-branes should shed more
light on the structure of D-branes and might hopefully give some clues
to the reasons for the existence of dualities. 

\listrefs
\end